\documentclass[conference, 10pt]{IEEEtran}
\IEEEoverridecommandlockouts

\ifCLASSINFOpdf
\else
   \usepackage[dvips]{graphicx}
\fi
\usepackage{url}

\usepackage{textcomp}
\usepackage{xcolor}
\usepackage{amsfonts}
\usepackage{amsmath}
\usepackage{amssymb}
\usepackage[ansinew]{inputenc} 
\usepackage{xcolor}
\usepackage{mathtools}
\usepackage{graphicx}
\usepackage{booktabs}
\usepackage[font={small}]{caption}
\usepackage{subcaption}
\usepackage{import}
\usepackage{multirow}
\usepackage{cite}
\usepackage[export]{adjustbox}
\usepackage{breqn}
\usepackage{mathrsfs}
\usepackage{acronym}
\usepackage{setspace}
\usepackage{bm}
\usepackage{stackengine}
\usepackage{listings}
\usepackage[normalem]{ulem}
\usepackage{fancyhdr}

\def\BibTeX{{\rm B\kern-.05em{\sc i\kern-.025em b}\kern-.08em
		T\kern-.1667em\lower.7ex\hbox{E}\kern-.125emX}}

\graphicspath{{./figures/}}
\newcommand{\eq}[1]{Eq.~\eqref{#1}}
\newcommand{\fig}[1]{Fig.~\ref{#1}}
\newcommand{\tab}[1]{Tab.~\ref{#1}}
\newcommand{\secref}[1]{Section~\ref{#1}}

\newcommand\rev[1]{\textcolor{blue}{#1}}
\renewcommand\rev[1]{{#1}}  

\newcommand{\mytexttilde}{{\raise.17ex\hbox{$\scriptstyle\mathtt{\sim}$}}}

\begin{document}
\fancypagestyle{firststyle}
{
    \fancyhf{}
    \chead{This paper has been accepted at the 2022 IEEE International Workshop on Signal Processing Advances in Wireless Communications (SPAWC)}
   \renewcommand{\headrulewidth}{0pt} 
}

\title{Human Tracking with mmWave Radars: a Deep Learning Approach with Uncertainty Estimation}

\author{%
\IEEEauthorblockN{Jacopo Pegoraro$^\ddag$ (\textit{Graudate Student Member, IEEE}) and Michele Rossi (\textit{Senior Member, IEEE})\\
\thanks{$^\ddag$ Corresponding author email: \tt{pegoraroja@dei.unipd.it}}
\thanks{The authors are with the Department of Information Engineering, University of Padova, Italy.}
\thanks{This work has been supported by the EU MSCA ITN project MINTS ``Millimeter-wave networking and sensing for beyond-5G'' (Grant no.~861222) and by MIUR (Italian Ministry of Education, University and Research) through the ``Departments of Excellence'' initiative (Law 232/2016).}
}}

\maketitle
\thispagestyle{firststyle}

\begin{abstract} 
\mbox{mmWave} radars have recently gathered significant attention as a means to track human movement within indoor environments. Widely adopted Kalman filter tracking methods experience performance degradation when the underlying movement is highly \mbox{non-linear} or presents \mbox{long-term} temporal dependencies. As a solution, in this article we design a  \mbox{convolutional-recurrent} Neural Network (NN) that learns to accurately estimate the position and the velocity of the monitored subjects from high dimensional radar data. The NN is trained as a probabilistic model, utilizing a Gaussian negative \mbox{log-likelihood} loss function, obtaining explicit uncertainty estimates at its output, in the form of time-varying error covariance matrices. A thorough experimental assessment is conducted using a 77~GHz FMCW radar. The proposed architecture, besides allowing one to gauge the uncertainty in the tracking process, also leads to greatly improved performance against the best approaches from the literature, i.e., Kalman filtering, lowering the average error against the ground truth from 32.8 to 7.59~cm and from 56.8 to 14~cm/s in terms of position and velocity tracking, respectively.
\end{abstract}


\IEEEpeerreviewmaketitle

\section{Introduction}

\IEEEPARstart{I}{ndoor} human tracking with low power \mbox{millimeter-wave} (\mbox{mmWave}) radar sensors has been receiving considerable attention in the last few years, due to its wide applicability to the Internet of Things (IoT)~\cite{shah2019rf}.
The typical aim of these systems is to exploit the reflected signal from human subjects to infer their {\it state} in the physical space, e.g., their position and movement speed~\cite{knudde2017indoor,zhao2019mid, pegoraro2020multiperson, pegoraro2021real}.

\rev{In this paper, we address the limitations of widely used Bayesian tracking techniques, such as the extended Kalman filter (EKF), which require strong assumptions about the movement process, e.g., constant velocity. Despite being widely used in the literature \cite{zhao2019mid, pegoraro2021real, pegoraro2020multiperson}, these methods only work sufficiently well in practice because of the high frame rates of mmWave radar devices, but their capability of grasping the complexity of human movement is severely limited. 
In real environments, people often follow random and unpredictable trajectories, which do not match standard radar target movement models. This causes large predicted uncertainties when using model-based Bayesian filtering approaches, reflecting the intrinsic limitations of legacy models in human movement analysis.}
To resolve this, we advocate the use of a \mbox{model-free} and \mbox{end-to-end} deep learning approach. In addition, and to the best of our knowledge, we are the first to introduce to the radar field the concept of \textit{heteroscedastic}, i.e., sample-varying, uncertainty estimation for neural network (NN) architectures. Modeling the uncertainty in the state estimates allows obtaining an error covariance matrix associated with the NN predictions: recently, this has been successfully applied to computer vision problems~\cite{kendall2017uncertainties}. Note that such covariances are most valuable in indoor radar systems to increase the performance of processing blocks such as {\it data association}, in the case of \textit{(i)} multiple subjects being tracked concurrently, or \textit{(ii)} multiple radars with overlapping fields of view. The contributions of our work are:

\textbf{1)} We design a maximum-likelihood convolutional-recurrent neural network (ML-CRNN), based on gated recurrent units (GRU)~\cite{cho2014learning}. This NN outputs an estimate of the current subject state (position and velocity) given an arbitrarily long sequence of past radar observations, without making any assumptions on the underlying movement process. The \mbox{ML-CRNN} handles both the \textit{vision} part of the problem, processing the raw data, and the \mbox{non-linear} target \textit{tracking} part.

\textbf{2)} The proposed ML-CRNN outputs an heteroscedastic error covariance matrix paired with each state prediction, which weighs the confidence level of the state estimates. This is achieved by making the NN output the error covariance matrix of the state estimate, and training it as a probabilistic model via a Gaussian negative log-likelihood (NLL) loss function.

\textbf{3)} We design \mbox{ML-CRNN} for \textit{end-to-end} training (no pre-processing). While in the literature denoising and clustering phases are customary, \cite{knudde2017indoor,zhao2019mid, pegoraro2021real}, ML-CRNN sequentially processes {\it raw} range-Doppler/range-azimuth radar images.

Numerical results are obtained on our own experimental data, using a Frequency Modulated Continuous Wave (FMCW) INRAS RadarLog device working in the $77-81$~GHz band. The evaluation scenario is challenging and realistic, with furniture and other humans, in addition to the tracked subject.

\section{FMCW mmWave radar signal model}
\label{sec:radar}

A multiple-input multiple-output (MIMO) FMCW radar allows the joint estimation of the distance, the angular position and the radial velocity of the target(s) with respect to the radar device. To achieve this, the radar transmits sequences of chirp signals and measures the frequency shift of the reflection at its receiving antennas. Next, we provide a brief overview of the FMCW radar signal model, detailing the parameters that are used in this work. For a more comprehensive description, the reader may refer to~\cite{patole2017automotive, winkler2007range}.

We use an INRAS RadarLog FMCW radar with one transmitting antenna and \mbox{$\Gamma=16$} receiving antennas, organized as a linear array. The frequency of the transmitted chirp signal (TX) is linearly increased from a base value of \mbox{$f_o = 77$~GHz} to a maximum \mbox{$f_1 = 81$~GHz} in \mbox{$T_c = 180$~$\mu$s} (a \textit{sweep}). We define the bandwidth of the chirp as \mbox{$B = f_1 - f_o = 4$~GHz}. The chirps are transmitted every \mbox{$T_{\rm rep} = 250$ $\mu$s} in sequences of \mbox{$P = 256$} sweeps. For each of the $16$ antenna elements, a mixer combines the attenuated and delayed received signal (RX) with the transmitted one, generating the intermediate frequency (IF) signal. The IF signal is sampled along three different dimensions. First, \emph{fast time} sampling returns \mbox{$N=1024$} points from each chirp. For the \emph{slow time} (or Doppler) sampling, $P$ samples, one per chirp from adjacent chirps, are taken with period $T_{\rm rep}$. Finally, the \emph{spatial} sampling relates to the $\Gamma$ receiving channels, spaced apart by a distance $d$, and enables the localization of the targets in the physical space. A discrete Fourier transform (DFT) is applied along each sampling dimension and the square of its magnitude is computed to extract the power density for each frequency component. The resulting 3-dimensional signal is referred to as range-Doppler-azimuth (RDA) map, and the position of the power peaks along each axis can be associated with the radial distance, the angular position and the radial velocity of the subjects~\cite{patole2017automotive}.
The RDA maps are outputted by the radar at every \mbox{time-step}, with a rate of $15$~fps, and have a dimension of \mbox{$1024\times 64 \times 64$} points, due to the resolution used for the DFT along the fast time, angular and slow time dimensions respectively.

\section{Method}
We define the state of a human subject at a certain \mbox{time-step} $t$ as the vector \mbox{$\mathbf{x}_{t} = \left[x_t, y_t, v^x_t, v^y_t\right]^T \in \mathbb{R}^4$}, containing the Cartesian coordinates of the subject in the space, $x_t$ and $y_t$, and the velocity components $v^x_t$ and $v^y_t$. Our aim is to track the subject, namely, to sequentially estimate their current state across time, using a sequence of past and present observations of the system (\textit{filtering} problem). To this end, we design a recurrent NN that extracts information from a sequence of $T$ consecutive radar RDA maps, identified by index \mbox{$t=1, \dots, T$}, and that performs a regression task producing an estimate of the state, $\hat{\mathbf{x}}_{t}$. In contrast to typical regression approaches based on NNs, we wish to estimate not only the state of the subject, but also the corresponding error covariance, \mbox{$\mathbf{\Sigma}_{t} = E\left[\left(\mathbf{x}_{t} - \hat{\mathbf{x}}_{t}\right)\left(\mathbf{x}_{t} - \hat{\mathbf{x}}_{t}\right)^T\right]$}.

\subsection{Learning from raw data}

\begin{figure}[t!]
	\begin{center}   
		\subcaptionbox{RD image, $\mathbf{M}^{\rm RD}_t$.\label{fig:rd}}{\includegraphics[width=4.3cm]{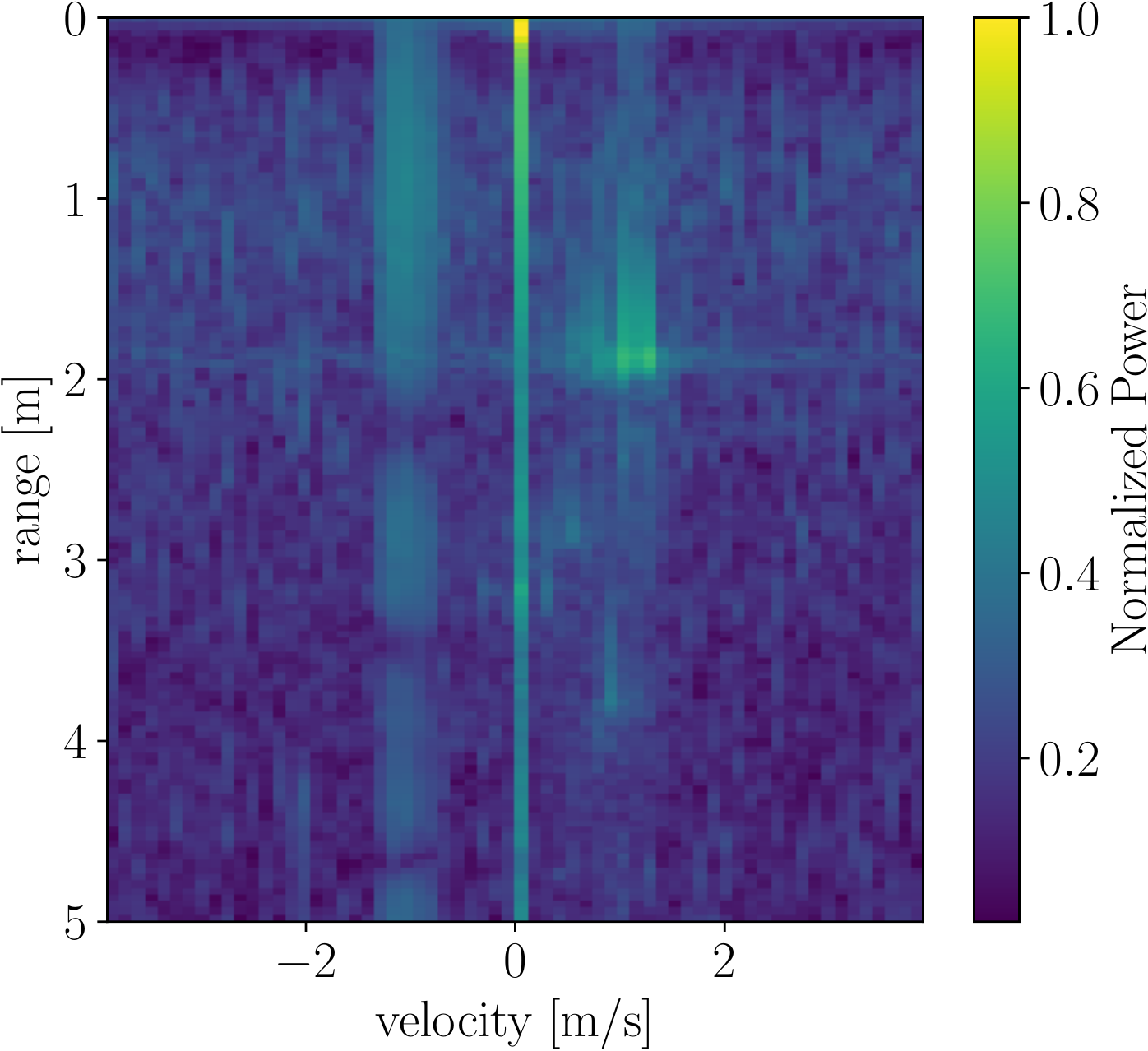}}
		\subcaptionbox{RA image, $\mathbf{M}^{\rm RA}_t$.\label{fig:ra}}{\includegraphics[width=4.3cm]{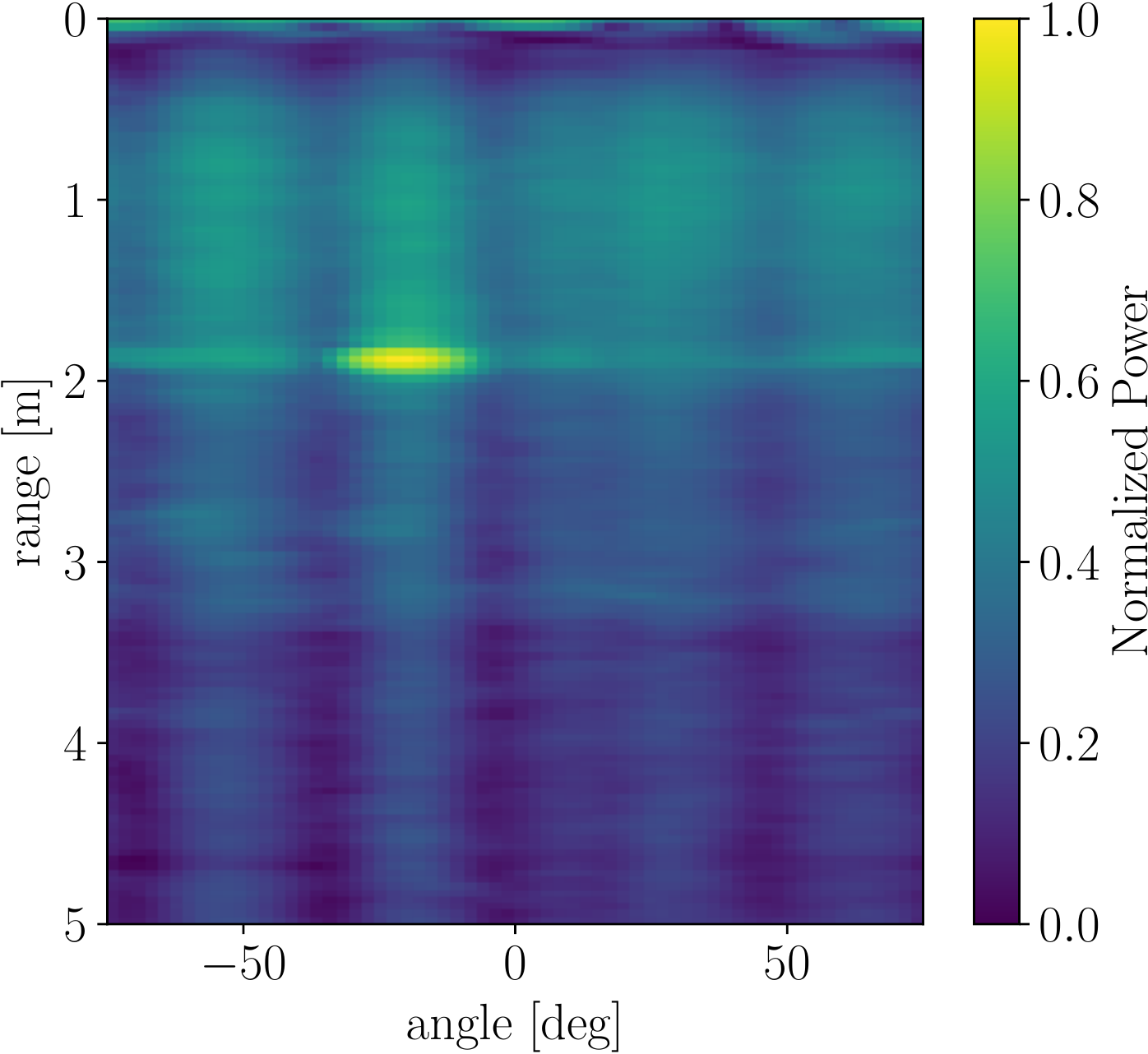}}
		\caption{Example RD and RA images. The target corresponds to the peak in received power around 2~m.}
		\label{fig:radar-maps}
	\end{center}
\end{figure}

Processing the raw RDA maps from the radar can be computationally very expensive given their size.
To mitigate this, we first select only the range interval of interest from the fast time dimension, i.e., the first $134$ points, that correspond to distances from $0$ to $5$~m. The resulting $134 \times 64 \times 64$ RDA map is then projected onto the range-Doppler (RD) plane by integrating along the azimuth dimension and onto the range-azimuth (RA) plane by integrating along the Doppler dimension. 
In this way, at each time-step $t$ we obtain a pair of $134\times 64$ images, denoted by $\mathbf{M}_t^{\rm RD}$ and $\mathbf{M}_t^{\rm RA}$, see \fig{fig:radar-maps}. Images are normalized so that pixels intensities lie in the interval $[0,1]$.

\subsection{Proposed neural network architecture}

\begin{figure}[t!]
	\begin{center}   
     \includegraphics[width=8.5cm]{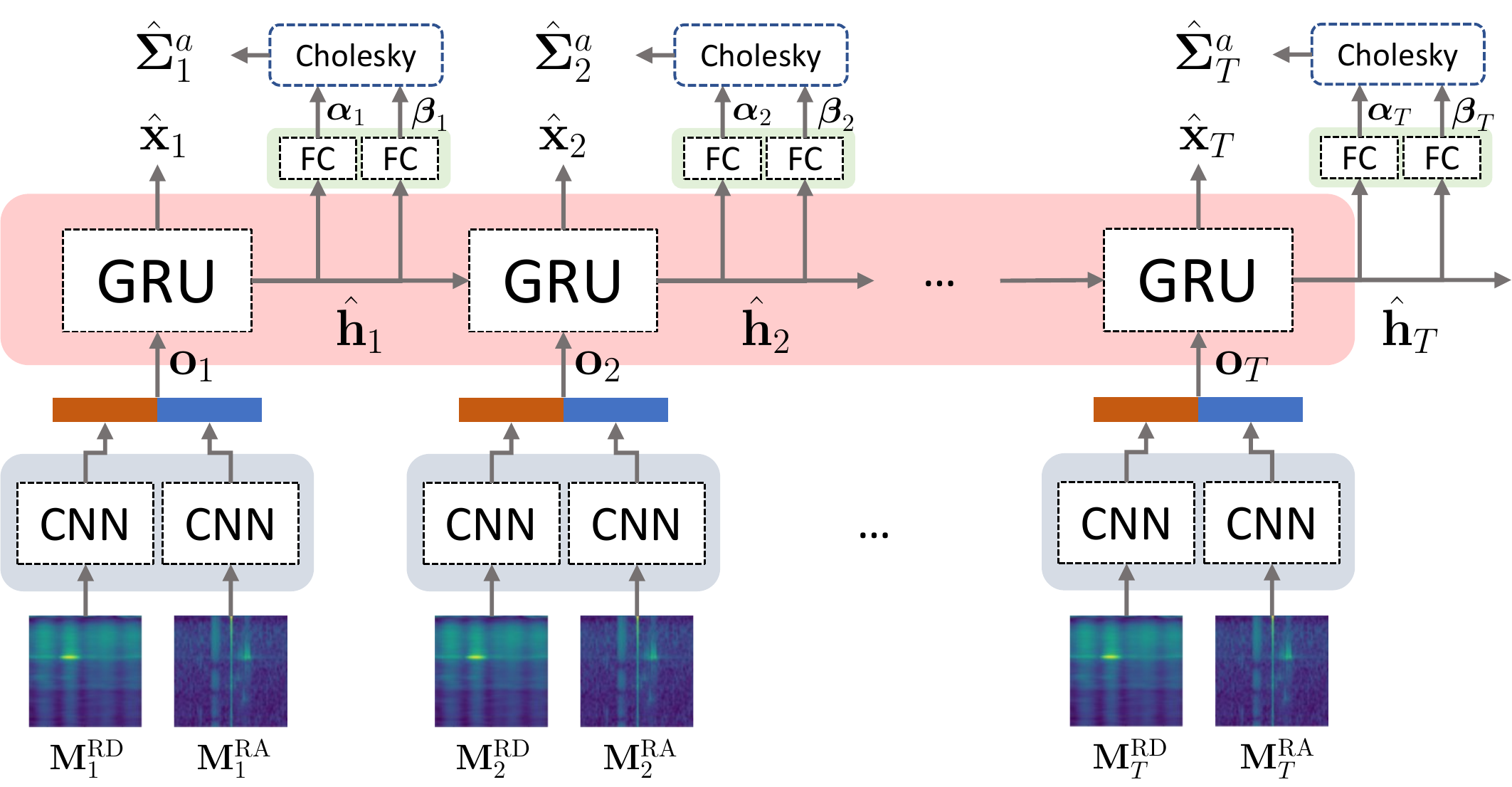}
	 \caption{Block diagram of the ML-CRNN architecture. }
		\label{fig:spawc-mlcrnn}
	\end{center}
\end{figure}

The proposed ML-CRNN combines convolutional layers operating on the single time-step with a recurrent structure based on GRU layers \cite{cho2014learning}, as shown in \fig{fig:spawc-mlcrnn}.
We can identify a convolutional and a recurrent block, which are connected together forming the ML-CRNN model and which are trained jointly via a common loss function (see \secref{sec:cov-est}).

The \textbf{convolutional block} is a convolutional neural network (CNN) that takes as input two images per time step, namely the RD and RA projections of the RDA map, \mbox{$\mathbf{M}_t^{\rm RD} \in \left[0,1\right]^{134\times 64}$} and \mbox{$\mathbf{M}_t^{\rm RA}\in \left[0,1\right]^{134\times 64}$}.
Given this input, the CNN learns a non-linear composite function $C$ that maps $\mathbf{M}_t^{\rm RD}$ and $\mathbf{M}_t^{\rm RA}$ onto a vector \mbox{$\mathbf{o}_t \in \mathbb{R}^{16}$}, called \textit{compressed} observation. The function is based on two parallel network branches that extract features from each input image separately, and that are then combined into a single output. Each branch, denoted by \mbox{$i \in \left\{{\rm RD}, {\rm RA}\right\}$}, is the composition of $L$ functions, which are the \textit{layers} of the CNN, $f^i_L\left(\dots f^i_1\left( \mathbf{M}_t^{i}\right)\right)$,
with \mbox{$L = 4$}.
Each layer $\ell$ computes the elementwise $\rm ELU$ activation function \cite{clevert2015fast} of the sum between the convolution of the input $X$ with $d_{\ell}$ \mbox{$3\times 3$} learned kernels, $\mathbf{K}_{\ell}^i$, and a bias parameter $b_{\ell}^i$: \mbox{$f_{\ell}^i\left( X\right) = {\rm ELU}\left(\mathbf{K}_{\ell}^i * X + b_{\ell}^i\right)$}.
The term $d_{\ell}$ represents the number of \textit{feature maps} of each layer and is equal to $4, 8, 16, 4$ for layer $\ell=1,2,3,4$, respectively. The kernels are applied using \textit{stride} \mbox{$2\times 2$}, that means they are shifted by two positions at each step of the convolution, resulting in a dimensionality reduction of a factor $2$ at each layer\footnote{Zero-padding  is applied to maintain the correct shape of the data.}. The output of each branch, is reshaped into a vector $\mathbf{y}^i_t $, and the two outputs are concatenated into $\mathbf{y}_t$. The final layer of the convolutional block processes $\mathbf{y}_t$ using a fully connected (FC) layer with \textit{dropout} probability $p=0.33$. Dropout refers to randomly setting to 0 the output of some NN nodes during training as a regularization method \cite{srivastava2014dropout}.
The FC layer applies the function \mbox{$\mathbf{o}_t =  {\rm ELU}\left(\mathbf{W}_{\rm fc}\mathbf{y}_t + \mathbf{b}_{\rm fc}\right) $}, with parameters $\mathbf{W}_{\rm fc}, \mathbf{b}_{\rm fc}$. Input radar images are processed in sequences of $T$ frames, applying in parallel the CNN block to each pair of RD and RA maps and obtaining a sequence of compressed observations, denoted by $\mathbf{o}_{1:T}$.

The \textbf{recurrent block} is a recurrent layer featuring GRU cells \cite{cho2014learning}. GRU cells maintain a hidden state across \mbox{time-steps}, processing it together with the current input vector to learn temporal dependencies in the input sequence (see \cite{cho2014learning} for the detailed description of a GRU cell).
The recurrent block takes as input $\mathbf{o}_{1:T}$ and outputs a sequence of estimates of the unobservable target states $\hat{\mathbf{x}}_{1:T}$ and the corresponding error covariance matrices $\hat{\mathbf{\Sigma}}_{1:T}$.
The hidden states are denoted by $\mathbf{h}_{1:T}$ and
have dimension $128$ each. 
At each time-step, they are further processed with $3$ FC output layers to compute the state estimate, $\hat{\mathbf{x}}_t \in \mathbb{R}^4$, and two vectors $\boldsymbol{\alpha}_t \in \mathbb{R}_+^4$ and $\boldsymbol{\beta}_t \in \mathbb{R}^6$, which are used to build the covariance matrix estimate $\hat{\mathbf{\Sigma}}_t$, as described in \secref{sec:cov-est}.
The expressions of the FC layers are the following
\begin{equation}\label{eq:state-est}
\hat{\mathbf{x}}_t  = \mathbf{W}_x\mathbf{h}_t + \mathbf{b}_x,
\end{equation}
\begin{equation}\label{eq:sigma-est}
\boldsymbol{\alpha}_t  = \exp\left(\mathbf{W}_{\alpha}\mathbf{h}_t + \mathbf{b}_{\alpha}\right),
\end{equation}
\begin{equation}\label{eq:rho-est}
\boldsymbol{\beta}_t  = \tanh \left(\mathbf{W}_{\beta}\mathbf{h}_t + \mathbf{b}_{\beta}\right),
\end{equation}
where we denoted by $\mathbf{W}_x, \mathbf{b}_x, \mathbf{W}_{\alpha}, \mathbf{b}_{\alpha}, \mathbf{W}_{\beta}, \mathbf{b}_{\beta}$ the learned weights and biases. In the recurrent block, \textit{recurrent dropout} is applied with probability \mbox{$p=0.33$} as described in~\cite{gal2016theoretically}. The ML-CRNN contains a total of $66730$ trainable parameters.

\subsection{Maximum likelihood state and covariance estimation} \label{sec:cov-est}

To model the uncertainty on the state estimates, we assume that the posterior distribution of the state given the observation sequence of length $T$ is Gaussian with mean $\hat{\mathbf{x}}_t$:
\mbox{$p(\mathbf{x}_t|\mathbf{M}^{\rm RD}_{1:T}, \mathbf{M}^{\rm RA}_{1:T}) \sim \mathcal{N}(\hat{\mathbf{x}}_t, \hat{\mathbf{\Sigma}}_t)$}.
Moreover, we let $\hat{\mathbf{\Sigma}}_t$ depend on the time-step, in order to reflect the variable uncertainty that affects radar measurements due to many factors, like the range-dependent power attenuation, the clutter distribution and the variability in the movement process.

The covariance matrix $\mathbf{\Sigma}_t$ must be symmetric and positive definite, and can be modeled as the sum of an aleatoric and an epistemic component~\cite{kendall2017uncertainties, russell2021multivariate}, as detailed next.

\textbf{1)} \textbf{Aleatoric covariance}, $\mathbf{\Sigma}_t^a$, is the uncertainty related to the {\it intrinsic noise} in the state evolution and measurement processes. It is estimated directly from the vectors $\boldsymbol{\alpha}_t, \boldsymbol{\beta}_t$ outputted by the ML-CRNN, see \eq{eq:sigma-est} and \eq{eq:rho-est}, using the Cholesky decomposition \mbox{$\hat{\mathbf{\Sigma}}_t^a = \mathbf{L}_t\mathbf{L}^T_t$} where $\mathbf{L}_t$ is a lower triangular matrix \cite{watkins2004fundamentals}.
To ensure that the covariance matrix is positive semi-definite it is sufficient that the diagonal elements of $\mathbf{L}_t$ are all non-negative. Vector $\boldsymbol{\alpha}_t$ is obtained using an exponential activation function, therefore its elements are positive and can be used as the diagonal elements of $\mathbf{L}_t$, namely \mbox{$[\mathbf{L}_{t}]_{i,i} = \alpha_{i, t}$}. 
The six off-diagonal elements of $\mathbf{L}_t$ correspond to the elements of vector $\boldsymbol{\beta}_t$, that are placed following an arbitrary (but consistent across iterations) order.

Once the aleatoric covariance is obtained applying the above transformations, 
a maximum-likelihood (ML) approach is used to jointly optimize the state and the covariance estimates, interpreting the training phase as fitting a probabilistic model. In particular, we use the negative log-likelihood of a multivariate Gaussian as the loss function of the ML-CRNN
\begin{equation}\label{eq:loss}
\ell(\mathbf{x}_t, \hat{\mathbf{x}}_t, \hat{\mathbf{\Sigma}}^a_t) = \left(\mathbf{x}_t - \hat{\mathbf{x}}_t\right)^T (\hat{\mathbf{\Sigma}}^{a}_t)^{-1}\left(\mathbf{x}_t - \hat{\mathbf{x}}_t\right) + \ln |\hat{\mathbf{\Sigma}}^a_t|,
\end{equation}
where both $\hat{\mathbf{x}}_t$ and $\hat{\mathbf{\Sigma}}^a_t$ are outputted by the network at each time-step. The total loss on the sequence of $T$ frames is computed as \mbox{$\mathcal{L} = \sum_{t=1}^T\ell(\mathbf{x}_t, \hat{\mathbf{x}}_t, \hat{\mathbf{\Sigma}}^a_t)/T$}. Training the network by minimizing \eq{eq:loss} amounts to maximizing the likelihood that the predicted state and covariance actually represent the parameters of a Gaussian probabilistic model.

\textbf{2)} \textbf{Epistemic covariance}, $\mathbf{\Sigma}_t^e$, is due to the {\it uncertainty in the prediction} made by the deep learning model. It can be estimated using Monte-Carlo (MC) dropout~\cite{gal2016dropout}. This method consists in applying the dropout procedure during inference, making the NN output random even for a fixed input. MC dropout uses the last NN available at time $t$, and is not part of the NN parameter learning process. 

\begin{figure}[t!]
	\begin{center}   
     \includegraphics[width=8.5cm]{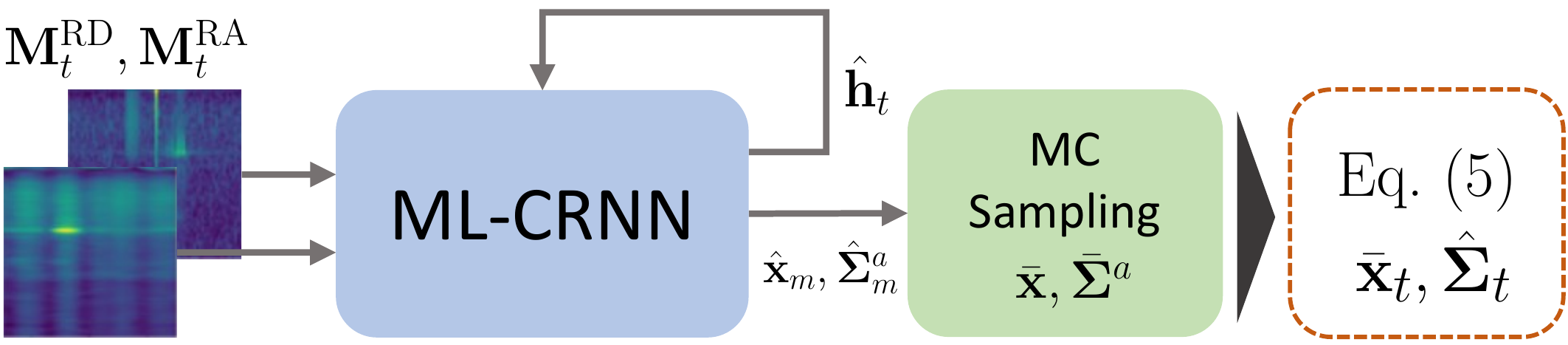}
	 \caption{Block diagram of the ML state and covariance estimation.}
		\label{fig:spawc-sys}
	\end{center}
\end{figure}

\textbf{Total variance}. The ML-CRNN model at time $t$ can be run $M$ times for each input with MC dropout. In this way, $M$ different state and covariance samples are obtained for the same input. The time index is dropped here for convenience, as we operate within a single time-step. We respectively denote by $\hat{\mathbf{x}}_m$ and $\hat{\mathbf{\Sigma}}_m^a$, $m = 1, \dots, M$, the state and covariance predictions outputted by the NN, while we refer to their empirical averages over the $M$ samples as $\bar{\mathbf{x}} = \sum_m \hat{\mathbf{x}}_m / M$ and $\bar{\mathbf{\Sigma}}^a = \sum_m \hat{\mathbf{\Sigma}}_m^a / M$. Using the sample covariance estimator for $\mathbf{\Sigma}^e$, and $\bar{\mathbf{\Sigma}}^a$ as the sample mean estimate of $\mathbf{\Sigma}^a$, the total error covariance matrix can be expressed as \cite{russell2021multivariate}
\begin{multline}\label{eq:tot-cov}
\hat{\mathbf{\Sigma}} \approx \underbrace{\frac{1}{M}\sum_{m=1}^M (\hat{\mathbf{x}}_m - \bar{\mathbf{x}})(\hat{\mathbf{x}}_m - \bar{\mathbf{x}})^T}_{\textrm{epistemic}} + \underbrace{\bar{\mathbf{\Sigma}}^a}_{\textrm{aleatoric}}.\\
= \frac{1}{M}\sum_{m=1}^M \hat{\mathbf{x}}_m \hat{\mathbf{x}}_m^T -\bar{\mathbf{x}}\bar{\mathbf{x}}^T + \bar{\mathbf{\Sigma}}^a.\\
\end{multline}
$\bar{\mathbf{x}}$ is typically more precise than a single sample from the MC dropout, so the final system outputs at step $t$ are the averaged estimate \mbox{$\hat{\mathbf{x}}_t= \bar{\mathbf{x}}_t$} and its covariance $\hat{\mathbf{\Sigma}}_t$ from \eq{eq:tot-cov}. The complete procedure is summarized in \fig{fig:spawc-sys}

\section{Results}
\begin{figure*}[t!]
	\begin{center}   
		\centering
		\subcaptionbox{\label{fig:nll}}{\includegraphics[width=6.1cm]{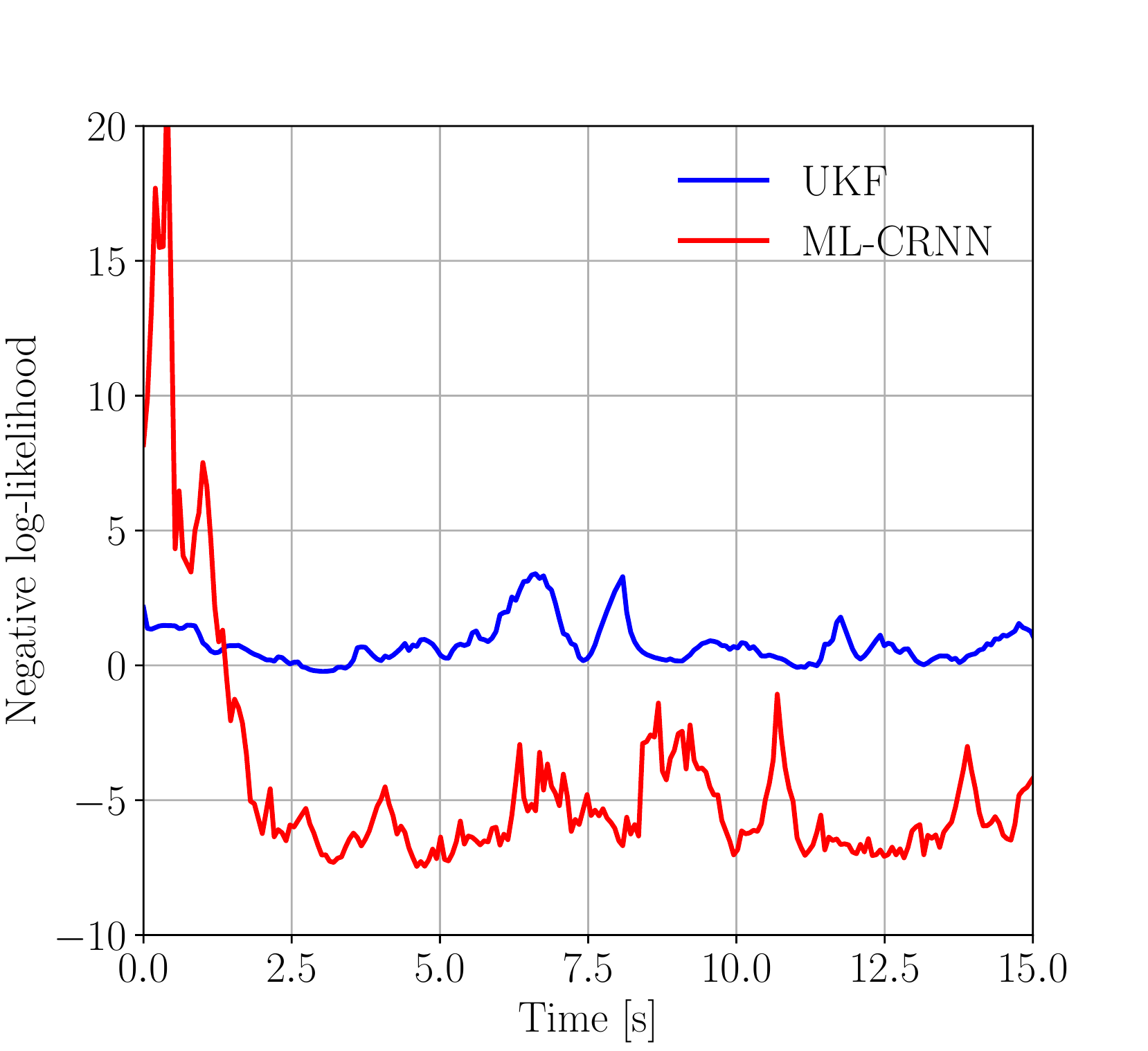}}
		\subcaptionbox{\label{fig:std-comp}}{\includegraphics[width=5.9cm]{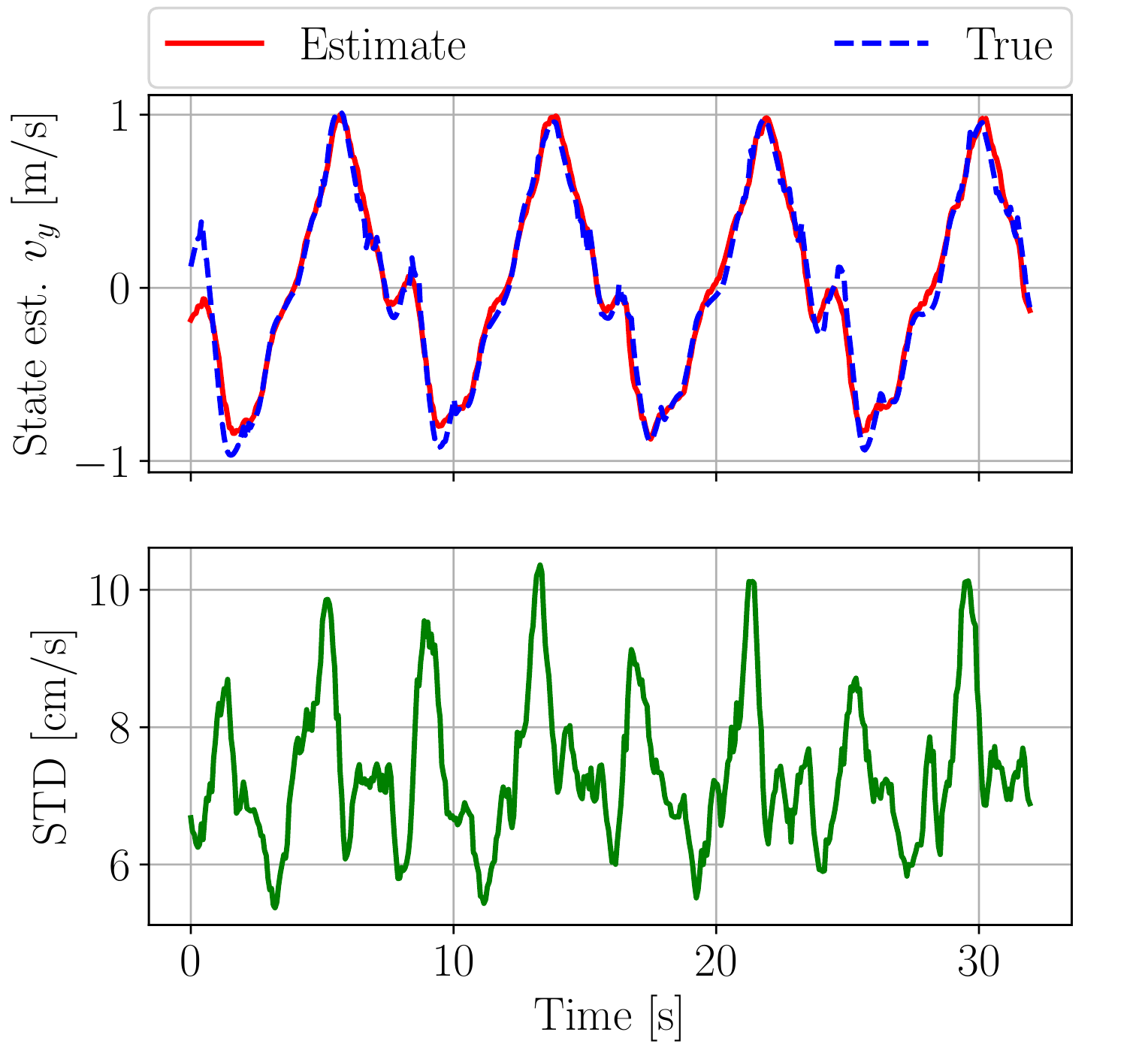}}
		\subcaptionbox{\label{fig:cal}}{\includegraphics[width=5.8cm]{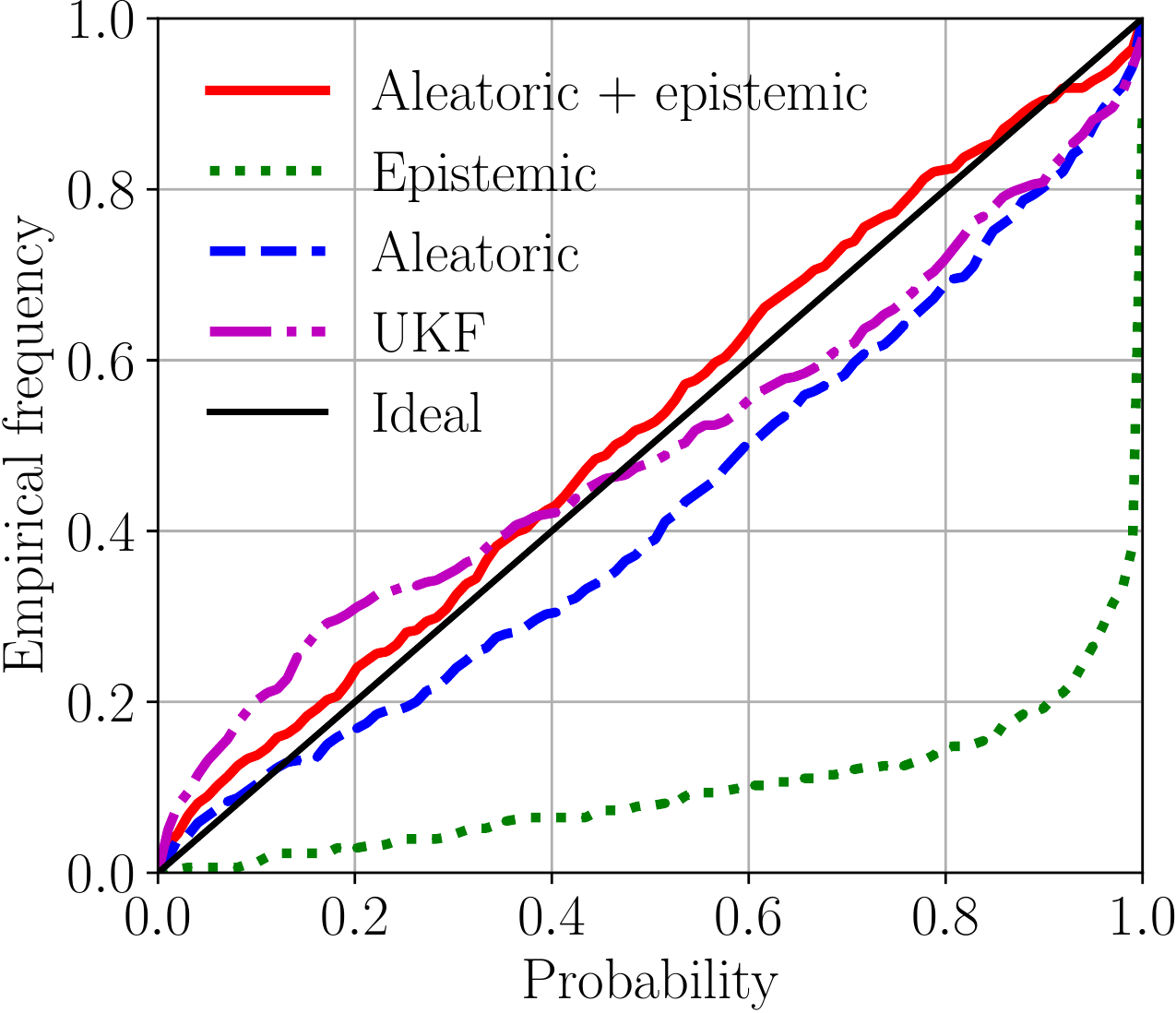}}
		\caption{(a) Negative log-likelihood (NLL) on test data for ML-CRNN and UKF. (b) Estimate of $v_y$ (top) and its predicted standard deviation (bottom). (c) Comparison between the empirical distribution of the squared Mahalanobis distances and the theoretical one.}
		\label{fig:results}
	\end{center}
\end{figure*}

The proposed NN architecture is trained and tested on experimental measurements taken in a $8$~m~$\times~4$~m research laboratory equipped with a motion tracking system with six \mbox{infra-red} cameras. A total of $10$~minutes of training data and $1$~minute of test data were collected for a single subject moving inside a $4$~m~$\times~2$~m rectangle, i.e., the working area of the motion tracking system. These measurements were taken in realistic conditions, with furniture and other people inside the room, but outside the tracking area. This makes the radar images highly cluttered, with bursts of frames in which the target subject is undetectable. Ground truth data is concurrently acquired by the motion tracking system, which was \mbox{time-synchronized} with the radar.

We trained the ML-CRNN using the loss in \eq{eq:loss} and the Adam optimizer \cite{kingma2014adam}, using a subset of the training data as the validation set. Training was stopped when the loss reached convergence on the validation set, and the epistemic covariance on the training set became negligible. The length of the temporal sequences that are fed to the NN \textit{during training} is \mbox{$T=10$} ($0.667$~s). During the evaluation, instead, the predictions are obtained by inputting a new radar frame in the ML-CRNN as soon as it becomes available. The ML-CRNN then uses the hidden state, $\mathbf{h}_t$, and the current input to compute the prediction, similarly to how Bayesian filtering methods operate. The following metrics were used to evaluate the tracking error: \textit{(i)}~root mean square error (RMSE) and \textit{(ii)}~localization error outage, \mbox{${\rm LEO}(0.2) = {\rm Prob}\left(||\mathbf{x} - \hat{\mathbf{x}}|| > 20 \mbox{ cm}\right)$}.
\subsection{Performance}
\begin{table}[t!]
	\small 
	\begin{center}
		\begin{tabular}{lccc}
			\toprule
			& \multicolumn{2}{c}{{\bf Position}} & {\bf Velocity} \\
			\cmidrule(lr){2-4}
			\textbf{Method} &  RMSE [cm] & LEO$\left(0.2\right)$ [\%] & RMSE [cm/s]\\
			\midrule
			UKF & 32.8 & 47.1 & 56.8\\
			MSE-CRNN & 12.8  & 7.30 & 20.1\\
			ML-CRNN & 7.59  & 0.62 & 14.0\\
			\bottomrule
		\end{tabular} 		
	\end{center}
	\caption{Tracking error of UKF and the proposed CRNN network with ML (ML-CRNN) and MSE (MSE-S2S) loss criteria. \label{tab:comp}}
\end{table}
\textbf{Tracking --} In \tab{tab:comp}, we show the results obtained by the proposed CRNN model with the ML loss of \eq{eq:loss} (ML-CRNN), compared to the same model trained with standard MSE loss (MSE-CRNN), i.e., without the covariance estimation, and to an unscented Kalman filter (UKF), which is a widely adopted Bayesian filtering method for estimating the posterior state distribution in non-linear radar tracking~\cite{julier1997new}. The parameters of the UKF have been optimized via grid search on the same training dataset used for the ML-CRNN.
Both CRNN methods are clearly superior in tracking accuracy to the UKF, showing an RMSE reduction of about $0.25$~m in the location accuracy and $0.4$~m/s in the velocity estimation. Note that the UKF cannot perform tracking from high-dimensional raw data: denoising and clustering are needed to transform the RDA maps into vectors containing the range and angular position of the target. For this purpose, we implemented the clustering method used in \cite{pegoraro2020multiperson, pegoraro2021real, zhao2019mid}, based on the DBSCAN~\cite{ester1996density} algorithm.

A further important aspect is the effect of NN training by using the ML loss: in addition to getting an estimate of the prediction uncertainty, we also observed a considerable improvement of the tracking accuracy. The ML-CRNN learning process is indeed less affected by outliers in the radar measurements due to its probabilistic nature, assigning low importance to unlikely observations. 


\textbf{Uncertainty estimation --} To gauge the quality of the obtained uncertainty estimates, we first compare the NLL from ML-CRNN against that of UKF, see \fig{fig:nll}, using $M=25$ MC samples in \eq{eq:tot-cov}. Note that, in practice, the value of $M$ can be tuned to trade off between computational complexity and quality of the resulting uncertainty estimation. We notice that ML-CRNN is slower to converge, due to its long-term dependency on past inputs, (\mbox{$T=10$} time-steps), but achieves much better performance (smaller NLL) after the initial transient, showing its superior capability of capturing the underlying human movement model.

In \fig{fig:std-comp}, we show the relation between the predicted uncertainty and the position estimates for ML-CRNN, focusing on the $y$ component of the velocity, $v_y$. We see that the uncertainty (standard deviation of $v_y$, bottom) shows a positive peak when $v_y$ changes rapidly (top).


A further way to investigate the quality of the covariance is to compare the empirical frequency of the squared Mahalanobis distance, $\xi_t = \left(\mathbf{x}_t - \hat{\mathbf{x}}_t\right)^T (\hat{\mathbf{\Sigma}}_t)^{-1}\left(\mathbf{x}_t - \hat{\mathbf{x}}_t\right)$, on the test measurements, with its theoretical probability distribution. 
Due to the Gaussian posterior probability assumption for the state, it can be shown that $\xi_t$ should follow a $\chi^2$ distribution with 4 degrees of freedom (the state dimension)~\cite{bar2009probabilistic}. 

In \fig{fig:cal}, we plot a comparison between the empirical frequency and the theoretical value of the probability distribution of $\xi_t$.  An ideal calibration of the uncertainty would yield a perfect match between the two, as in the black line. From our experiment, we see that a clear improvement is obtained with ML-CRNN by using both the aleatoric and the epistemic components of the covariance, over using either of them in isolation. In particular, only using the epistemic component leads to severely underestimating the uncertainty, because it neglects the intrinsic variability of the movement process of the target. 
On the other hand, the UKF shows inferior calibration quality, which denotes the limitations of the underlying movement model. Quantitatively, the calibration mean-squared errors between the ideal case (perfect calibration) and the predicted uncertainty are $9\cdot 10^{-4}$ and $5\cdot 10^{-3}$ for the ML-CRNN ad the UKF, respectively.

\section{Conclusions}
In this paper, we proposed a convolutional recurrent neural network to track human movement in indoor spaces by means of a mmWave MIMO FMCW radar. Our model estimates position and velocity of the subjects from raw radar data with  superior accuracy with respect to \mbox{state-of-the-art} techniques, and without requiring any assumptions on the movement evolution process. The proposed neural network is trained as a probabilistic model using a maximum-likelihood loss function, obtaining explicit {\it uncertainty estimates} at its output, in the form of a time-varying error covariance matrices. This, besides allowing one to gauge the uncertainty in the tracking process, also leads to greatly improved performance against the best approaches from the literature, i.e., the unscented Kalman filter, lowering the average tracking error from $32.8$ to $7.59$~cm and from $56.8$ to $14$~cm/s in terms of position and velocity, respectively.

Future research work includes the integration of deep learning models for object detection and recognition in the ML-CRNN. This would allow simultaneously detecting multiple targets, obtaining a probability distribution of their position and recognizing the target nature, e.g., person, pet, vehicle, etc. 

\bibliography{biblio}
\bibliographystyle{ieeetr}

\end{document}